# Persistent Magnetism in Silver-doped BaFe$_2$As$_2$ Crystals

Li Li,[1] Huibo Cao,[2] David S. Parker,[1] Stephen J. Kuhn,[1,3] and Athena S. Sefat[1]

[1] *Materials Science & Technology Division, Oak Ridge National Laboratory, Oak Ridge, TN 37831*
[2] *Quantum Condensed Matter Division, Oak Ridge National Laboratory, Oak Ridge, TN 37831*
[3] *Department of Physics, University of Notre Dame, Notre Dame, IN 46556*

**Abstract**

We investigate the thermodynamic and transport properties of silver-substituted BaFe$_2$As$_2$ (122) crystals, up to ~4.5%. Similar to other transition-metal substitutions in 122, Ag diminishes the antiferromagnetic ($T_N$) and structural ($T_S$) transition temperatures, but unlike other electron-doped 122s, $T_N$ and $T_S$ coincide without splitting. Although magnetism drops precipitously to $T_N = 84$ K at doping $x = 0.029$, it only weakly changes above this $x$, settling at $T_N = 80$ K at $x = 0.045$. Compared to this persistent magnetism in Ag-122, doping other group 11 elements of either Cu or Au in 122 diminished $T_N$ and induced superconductivity near $T_C = 2$ K at $x = 0.044$ or 0.031, respectively. Ag-122 crystals show reflective surfaces with surprising thicker cross sections for $x \geq 0.019$, the appearance that is in contrast to the typical thin stacked layered feature seen in all other flux-grown x122 and lower Ag-122. This physical trait may be a manifest of intrinsic weak changes in *c*-lattice and $T_N$. Our theoretical calculations suggest that Ag doping produces strong electronic scattering and yet a relatively small disruption of the magnetic state, both of which preclude superconductivity in this system.

## I. INTRODUCTION

High-temperature superconductivity (HTS) is the most mysterious and elusive property in condensed-matter physics found in two Cu- and Fe-based material families.[1,2] The iron-based superconductors (FeSC) share some common features with the cuprates,[3,4] most importantly that the superconducting state is triggered by chemical doping (or pressurizing) of an antiferromagnetic (AF) 'parent' material.[3–6] On the other hand, the parents of FeSC have Fermi surfaces that are sensitive to small changes in composition,[7–9] and that certain applications of pressure[6] or chemical substitutions[10] in the Fe-plane can instigate superconductivity. In FeSC (see refs.[4,11–22] for several reviews), there is a highly complex interplay of factors such as competition between magnetism and superconductivity, close proximity of lattice distortion to the $T_N$ and the associated nematicity, orbital ordering, local moment, twinning, disorder and chemical clustering. Despite the rich physical chemistry that FeSC offers,[23] and the vast experimental and theoretical work seen in review manuscripts,[24–29] many things about them remain unpredictable (*e.g.*, chemistry doping trends, reasons for HTS and the particular $T_N$ and $T_C$ values).

BaFe$_2$As$_2$ transitions from tetragonal (*I*4/*mmm*) non-magnetic state into orthorhombic (*Fmmm*) striped-AF phase below $T_S = T_N = 132\text{-}134$ K (flux-grown crystals).[30,31] For 122, transition-metal doping with either holes (*e.g.*, 3*d* Cr, Mn; 4*d* Mo)[32–35] or electrons (*e.g.*, 3*d* Co, Ni; 4*d* Rh, Pd)[10,36–38] suppresses AF, but only electron dopants can instigate superconductivity. The reason for this is not exactly solved, especially since the dopants can be very low in concentration (< 8%). However, a couple of trends are noted in literature. One, electron doping using 3d or 4d in the same group (Co and Rh, or Ni and Pd) gives overlapping temperature-composition ($T$-$x$) phase diagrams.[31] However, this breaks for 5d, where Pt-122 gives much wider $x$ superconducting region ($x \approx 0.01$ to $0.11$),[39,40] while Ir-122 has highest $T_C$ (28 K) for $x = 0.15$.[41] Two, for electron doping using Co,[42,43] Ni,[44] Rh,[38,45] Pd,[38] Ir,[46] or Cu,[37] the $T_S$ and $T_N$ decrease while they decouple ($T_S > T_N$) with increasing $x$, eventually giving superconductivity. In contrast, both transitions happen at identical temperatures for all 122 hole-doping cases of Cr,[47] Mn,[48,49] and Mo,[33] and no superconductivity emerges in them. This manuscript is the first case study of the effects of 4d Ag-doping in BaFe$_2$As$_2$ (Ag-122) and we construct and compare $T$-$x$ phase diagram to that of the closely related 3d Cu-122.

In this paper, we provide experimental evidence for the properties of Ag-122 crystals using the variety of temperature-dependent magnetization, resistivity, heat capacity, Hall effect, and neutron diffraction measurements. Comparing 4d Ag-122 to 3d Cu- and 5d Au- dopants of the same group, similar suppression of $T_N$ is seen up to $x \sim 0.02$. But while Cu- and Au-122 give superconductivity at $\sim$ 2 K at $x = 0.044$[37] and $0.031$,[50] respectively, Ag remains magnetic up to $x = 0.045$. Our results show that Ag substitution gives coupled $T_S/T_N$ feature similar to hole-doped non-superconducting 122s. Our density of state calculations show that Ag creates essentially a band separate from the rest of the 122 electronic states, indicative of intense electronic scattering, different from that seen for total superconductors such as Co- or Cu-122s.

**Table 1** For Ba(Fe$_{1-x}$Ag$_x$)$_2$As$_2$, loading reaction ratio (Ag:FeAs), silver amount found from EDS ($x$); room-temperature lattice parameters refined from X-ray diffraction data; summary of transition temperatures inferred from resistivity, magnetization and specific heat measurements and neutron results.

| Ag : FeAs | $x$ | $c$ (Å) | $a$ (Å) | $T_N$, $T_s$ (K) | | | |
|---|---|---|---|---|---|---|---|
| | | | | ($dR/dT$) | ($d\chi/dT$) | ($dC/dT$) | (*neutrons*) |
| 0 : 5 | 0 | 13.0151(3) | 3.9619(2) | 132 | 132 | 132 | 133 |
| 0.1 : 5 | 0.005 | 13.0236(6) | 3.9647(2) | 120 | 120 | 120 | 125 |
| 0.2 : 5 | 0.009 | 13.0278(10) | 3.9668(4) | 114 | 108 | 111 | - |
| 0.3 : 5 | 0.016 | 13.0293(10) | 3.9685(3) | 106 | 106 | 106 | 105 |
| 0.4 : 5 | 0.019 | 13.0294(10) | 3.9714(3) | 101 | 96 | 100 | - |
| 0.5 : 5 | 0.026 | 13.0372(10) | 3.9736(3) | 90 | 92 | 93 | 93 |
| 0.6 : 5 | 0.029 | 13.0389(6) | 3.9749(2) | 76 | 86 | - | 84 |
| 0.7 : 5 | 0.035 | 13.0367(8) | 3.9775(2) | 80 | 82 | - | 80 |
| 0.9 : 5 | 0.040 | 13.0363(4) | 3.9789(3) | 79 | 79 | - | - |
| 1.1 : 5 | 0.045 | 13.0398(6) | 3.9801(1) | 75 | 79 | - | 80 |

## II. EXPERIMENTAL RESULTS AND DISCUSSION

Single crystals of Ag-doped $BaFe_2As_2$ were grown out of FeAs self-flux technique.[23] To produce a range of dopant concentrations, small barium chunks, silver powder, and FeAs powder were combined according to various loading ratios of Ba:Ag:FeAs = 1:$x$:5 (listed in Table 1) in a glove box, and each placed in an alumina crucible. A second catch crucible containing quartz wool was placed on top of each growth crucible, and both were sealed inside a silica tube under ~1/3 atm argon gas. Each reaction was heated for ~24 h at 1180 °C, and then cooled at a rate of 1 to 2°C/h, followed by a decanting of the flux around 1050 °C. The crystals were flat with dimensions of ~ $6 \times 4 \times 0.1$ mm$^3$ or smaller. Similar to 122,[30] the crystals of Ag-122 formed with the [001] direction perpendicular to the flat faces. Attempts for higher Ag contents were unsuccessful and only led to inhomogeneous phase. The chemical composition of each crystal batch was measured with a Hitachi S3400 scanning electron microscope operating at 20 kV; energy-dispersive x-ray spectroscopy (EDS) indicated that significantly less Ag is chemically substituted in the 122 structure than put in solution. Three spots (~ 80 μm) were checked and averaged on each random crystalline piece; the crystals had the same composition within each batch within error; no impurity phases or inclusions were detected. The samples are denoted by measured EDS $x$ values (each $x$ with a relative uncertainty of 5%) in $Ba(Fe_{1-x}Ag_x)_2As_2$ throughout this paper (Table 1). Although Ag-122 crystals have similar flat crystalline faces along the $ab$ plane (marked by yellow arrows in Fig1.a), they exhibit two different crystalline features in cross section (marked by green arrows in Fig1.b). For $0 \leq x < 0.019$, the cross section clearly shows a stacked layered feature, while for $0.019 < x \leq 0.045$ the cross section displays a uniform reflective surface with non-layered feature. We have not seen such thicker cross-sectional features in other doped 122 systems.

Bulk phase purity of Ag-122 crystals was checked by collecting data on an X'Pert PRO MPD X-ray powder diffractometer using monochromatic Cu $K_{\alpha 1}$ radiation in the 10-70° 2$\theta$ range, on ground crystals, each weighing ~ 30 mg collectively. Lattice parameters were refined from full-pattern refinements using X'Pert HighScore Plus software. The Bragg reflections were indexed using the tetragonal $ThCr_2Si_2$ tetragonal structure ($I4/mmm$), with no impurity phases. The refined lattice constants are listed in Table 1; Fig. 1c plots $a$- and $c$- lattice parameters as a function of $x$ for Ag-122. With Ag doping, the lattice parameter $c$ firstly increases at $x < 0.01$, then exhibits a step-shape jump at $x = 0.019$ and remains nearly unchanged above; however, the lattice parameter $a$ keeps increasing linearly up to the doping limit $x = 0.045$, which confirms the effective substitution of Ag into system, reflected in cell volume expansion with $x$ (Fig. 1c, inset). Compared to 122, $a$ increases by 0.39% for Ag-122 at $x = 0.035$, larger change than the 0.22% for Cu-122 at same $x$;[37] Moreover, Ag-doping also expands $c$, in contrast with Cu doping. For Co-122, both $a$ and $c$ decrease monotonically. The different doping effects on 122 lattice should be related to ionic radii variations, assuming +2 oxidation states, i.e. $Ag^{2+}$ (94 pm) > $Fe^{2+}$ (78 pm) > $Co^{2+}$ (74 pm) > $Cu^{2+}$ (73 pm).[51]

Magnetization measurements for $Ba(Fe_{1-x}Ag_x)_2As_2$ were performed in a Quantum Design (QD) Magnetic Property Measurement System (MPMS). For a temperature-sweep experiment, the sample was cooled to 2 K in zero field (zfc) and data were collected by warming from 2 to 300 K in an applied field of 1 Tesla. Fig. 2(a) and (b) present the magnetic-susceptibility results along $ab$- and $c$- crystallographic directions. For $BaFe_2As_2$, the susceptibility decreases approximately linearly with decreasing temperature, then drops abruptly below $T_N = T_S \approx 132$ K, reproducing the well-established behavior.[30,52] There is a small anisotropy as $\chi_{ab}$(300 K) = 0.92 $\times 10^{-3}$ cm$^3$/mole and $\chi_c$(300 K)= 0.68$\times 10^{-3}$ cm$^3$/mole. In the whole doping series, all the absolute values of $\chi_{ab}$ are larger than $\chi_c$ at room temperature. However, for all Ag-122 above ~ 150 K, the susceptibility data nearly display comparable linear dependence; they are neither Pauli nor Curie-Weiss like behavior, attributed to the multi-band nature of FeSCs and the spin-density-wave (SDW) nature of local and itinerant electrons.[53] For $0 \leq x \leq 0.026$, $\chi(T)$ displays similar temperature behavior, although the transition temperatures are reduced with $x$. For $x$ = 0.005, 0.009, 0.016, 0.019, and 0.026, $T_N$ values are inferred as ≈ 120 K, 108 K, 106 K, 96 K, and 92 K, respectively, using the $\chi$ derivative method as in Ref.[37]. The full list for these values is summarized in Table 1. For $x \geq 0.029$ crystals, the change in $T_N$ is small and the transitions are not as sharp as in the 122 parent (Fig.2 (a), inset). For Ag-122, the $T_N$ remains high and near 80 K at $x = 0.045$, in contrast to the fully suppressed AF and $T_C$ = 2 K in Cu-122 at $x = 0.044$, or $T_N < 50$ K and $T_C = 15$ K in Co-122 at $x = 0.047$.[54]

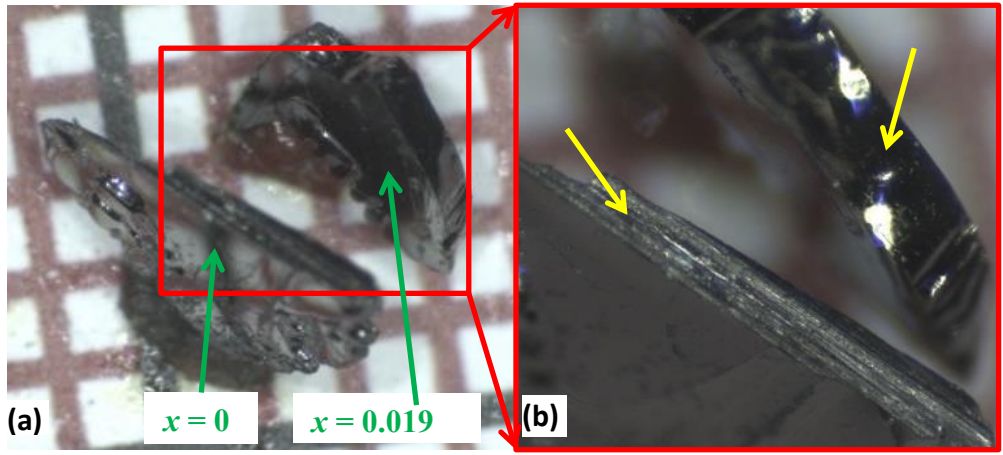

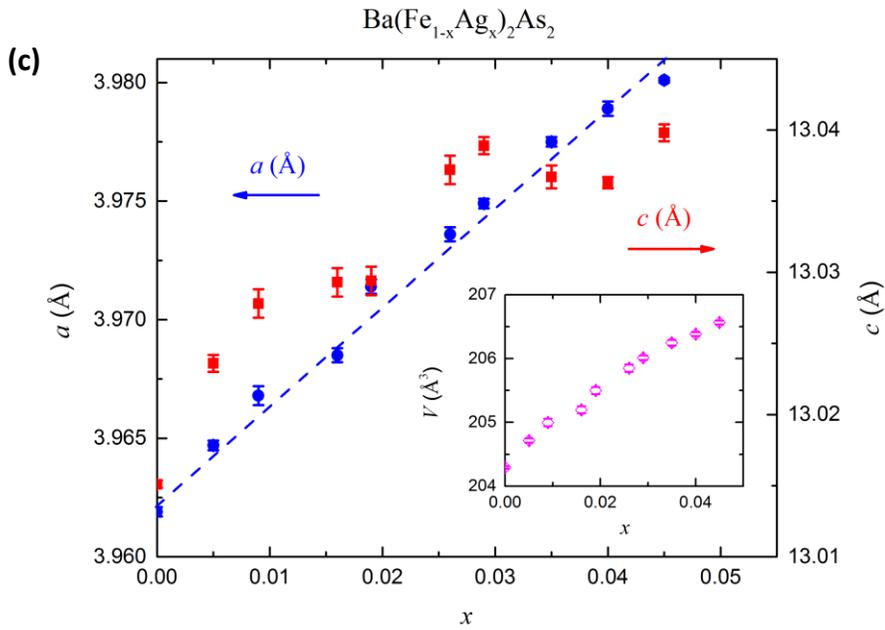

**Figure 1** (a) Typical single crystals for $x$ = 0 and 0.019. (b) Enlarged view for the cross-section of $x$ = 0 and 0.019 crystals. (c) Refined lattice parameters for $0 \leq x \leq 0.045$ in Ba(Fe$_{1-x}$Ag$_x$)$_2$As$_2$ series, inset is cell volume $V$ versus $x$.

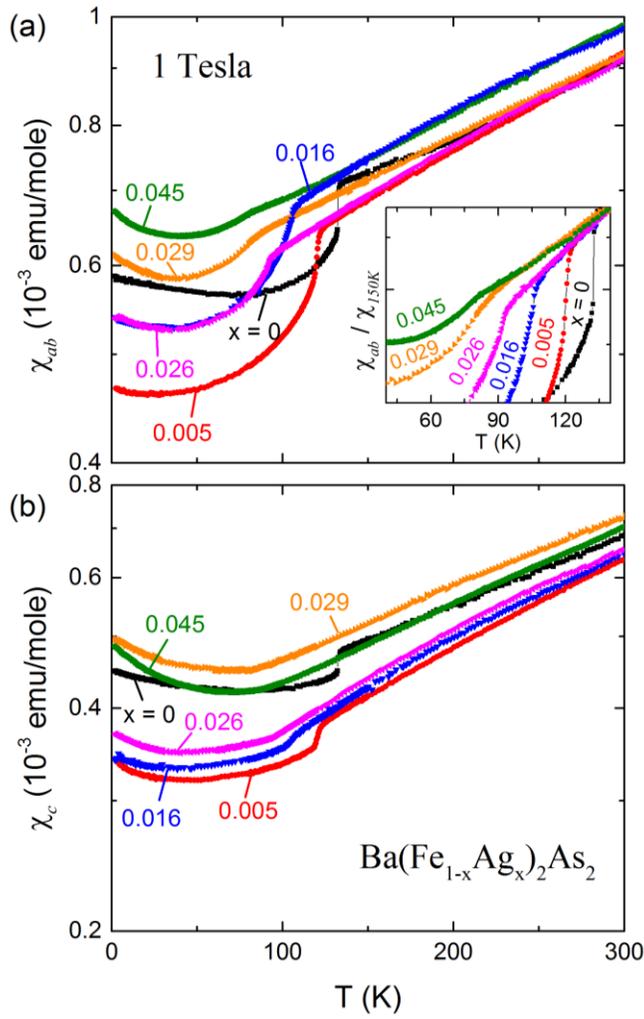

**Figure 2** For Ba(Fe$_{1-x}$Ag$_x$)$_2$As$_2$, temperature dependence of magnetic susceptibility for the range of 0 ≤ $x$ ≤ 0.045, (a) along $ab$-, and (b) $c$-lattice directions. Inset of (a) displays the normalized results along ab-direction in the temperature range of 40 K ⩽ T ⩽140 K.

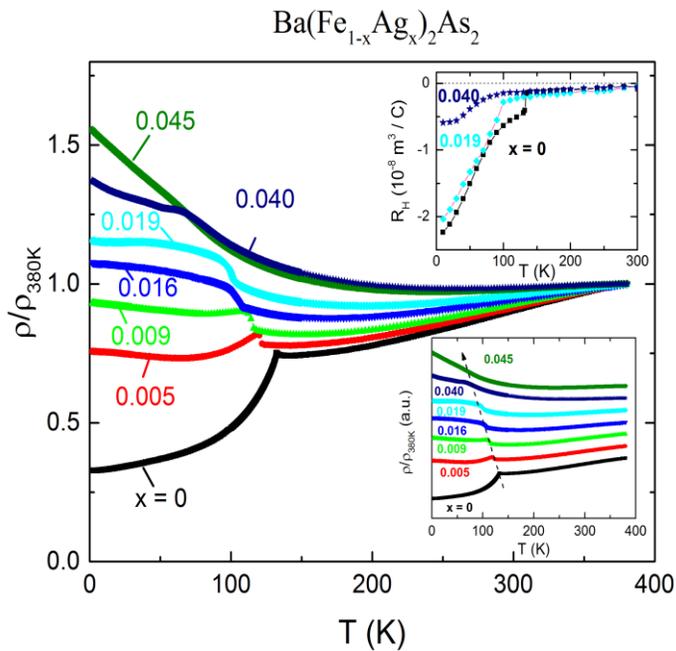

**Figure 3** For Ba(Fe$_{1-x}$Ag$_x$)$_2$As$_2$, temperature dependent resistivity for 0 ≤ x ≤ 0.045 normalized to 380 K the bottom inset has arbitrary ρ, the top inset shows Hall coefficient for x = 0, 0.019 and 0.040.

The electrical transport measurements were performed in a Quantum Design (QD) Physical Property Measurement System (PPMS). Electrical leads were attached to the crystals using Dupont 4929 silver paste and resistivity measured in the *ab* plane in the range of 1.8 to 380 K. The ρ values at 380 K ranged from 0.1 to 1.2 mΩ cm, although their absolute values may have suffered from the geometry factor estimations. Fig. 3 presents normalized $\rho/\rho_{380K}$; in inset, each *x* is shifted upward by 0.2 to clarify anomalies. Electrical resistivity for 122 is as expected, and the anomaly transition temperature $T^*$ (correspondent to $T_N$, $T_S$) is suppressed monotonically with increasing *x* similar to literature.[30,32–38] The resistivity of the parent Ba-122 exhibits an abrupt decrease below $T^*$. Despite the loss of carrier density, the decrease in the resistivity is due to a reconstruction of the Fermi surface in the orthorhombic striped-AF phase that generates high-mobility carriers dominating the charge transport.[55,56] For lightly Ag-doped composition of *x* = 0.005, the anomaly manifests an abrupt resistivity peak around 122 K, similar to that found in $Ba(Fe_{0.9923}Cu_{0.0077})_2As_2$,[37] followed by a decrease with cooling. For *x* = 0.009, the anomaly displays an increase around 112 K, followed by an almost flat resistivity dependence below. The resistivity for *x* ≥ 0.016 first decreases gently from 380 K, followed by upturns below 106 K for *x* = 0.016, 101 K for *x* = 0.019, and 79 K for *x* = 0.040. Such upturns below $T^*$ and continued increase of ρ with decreasing temperature are similar to what occurs in other electron-doped crystals.[10,36–38] The upturn reflects the loss of carriers as a partial SDW gap opens below $T_N$. At temperatures well below $T_N$, the increase in the mobility of the remaining carriers is not enough to overcome the lower carrier concentration and the resistivity continues to increase. For *x* = 0.045, the transition is too weak to be observed, similar to that seen in magnetic-susceptibility behavior. The inferred transition temperatures were extracted by the derivative of resistivity curve (*dρ/dT*) and are summarized in Table 1. In the whole doping series of Ag-122, no drop to zero resistivity is seen up to the chemical doping limit. This is in contrast to all other transition-metal electron-doped 122 such as in Co, Ni, Rh, Pd, Ir, Pt, Cu and Au, which give superconductivity in the comparable *x* regions.[10,36–41,54]

In order to gain more insight into the evolution of transport properties, the temperature dependence of hall coefficient ($R_H$) for *x* = 0, 0.019, and 0.040 is presented in Fig. 3, upper inset. The $R_H$ of pure $BaFe_2As_2$ is negative in the temperature region of 10-300 K, and shows a sharp decrease at the structural/magnetic transition near 132 K as reported before.[57] The values of $R_H$ for *x* > 0 are also negative between 10 and 300 K, with features at 100 K for *x* = 0.019, and ~80 K for *x* = 0.040, consistent with Fermi surface gapping scenario for $T_N$.[55,56] These anomalies are coincident with inferred transitions in $\chi(T)$ and $\rho(T)$. The overall change of Hall data for *x* = 0.019 and 0.040 are not as rapid as 122, which signify a weaker electronic structure change and potentially reduced magnetism. The widths of transitions for *x* = 0.040 is more broad than *x* = 0 and 0.019. The values of $R_H$ for *x* = 0.019 and 0.040 are less negative than that of parent 122 in the low temperature range. The negative sign of $R_H$ indicates that electrons give the dominant contribution to the charge transport in 122 and Ag-doped 122s, and further analysis is complicated by multiband nature of the system and the presence of both electron and hole bands at the Fermi level.

Specific heat data were measured also using a PPMS, shown in Fig. 4. For 122, a sharp transition is observed at 132 K, as expected, for overlapping $T_N$ and $T_S$. With Ag doping, the transition temperatures decrease monotonically, and the anomalies change from sharp peaks to broadened features. Such broadening feature is comparable to Cu-122,[37] and Au-122.[50] *dC/dT* is plotted in the left upper inset of Fig. 4, to show the transition temperatures. For *x* = 0.005, 0.009, 0.016, and 0.026, the anomalies occur at 120, 111, 105, and 92 K, respectively. There is no detectable transition in the *C(T)* curve for *x* = 0.045, which suggests its weak nature or highly disordered/strained crystals. The Sommerfeld-coefficient *γ* for all the compositions is estimated to be between 6 to 16 mJ·mol$^{-1}$·K$^{-2}$, as shown in bottom inset of Fig. 4. This weak change in *γ* with *x* is similar to that observed for $Ba(Fe_{1-x}Au_x)_2As_2$ and $Ba(Fe_{1-x}Mo_x)_2As_2$,[50,33] as may be expected for such low-doping levels. Moreover, the slightly doped Ba(Fe$_{1-x}$Co$_x$)$_2$As$_2$ (for $x$ =0.045, $\gamma$~14 mJ·mol$^{-1}$·K$^{-2}$) also has a weak change in $\gamma$.[58]

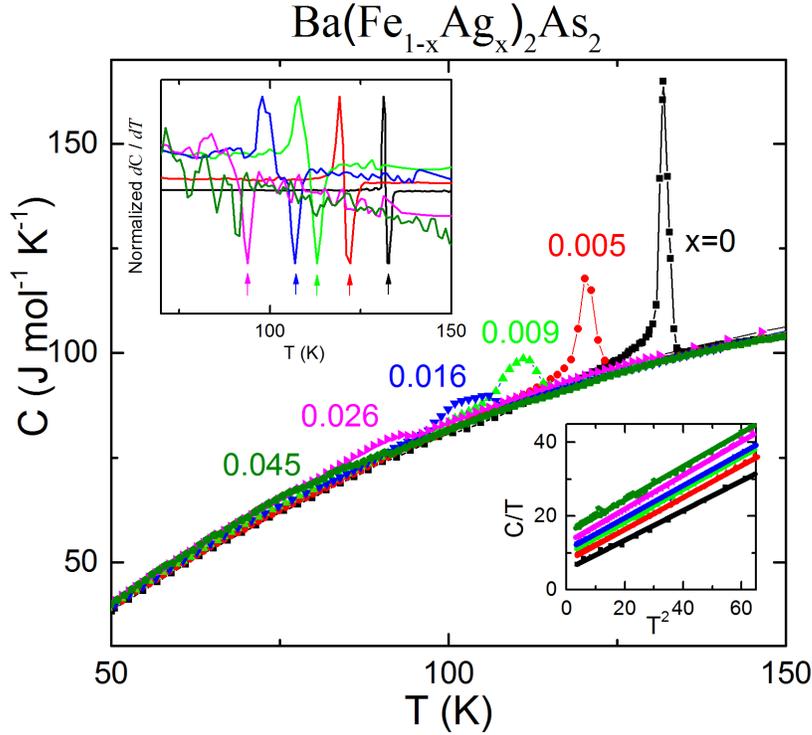

**Figure 4** For Ba(Fe$_{1-x}$Ag$_x$)$_2$As$_2$, heat capacity for 0 ≤ x ≤ 0.045 between 50 K and 150 K, the left top inset shows normalized *dC/dT* around the transitions, the right bottom inset displays the *C/T* versus T$^2$ at low temperature.

Single crystal neutron diffraction was performed using the four-circle diffractometer HB-3A at the High Flux Isotope Reactor (HFIR) at the Oak Ridge National Laboratory, to distinguish between the structural and magnetic transitions for $x$ = 0, 0.005, 0.029, 0.035 and 0.045. The neutron wavelength of 1.542 Å was used from a bent perfect Si-220 monochromator.[59] According to neutron diffraction data (Fig. 5), for 122 and as found before, there is a simultaneous structural and magnetic transition. In the magnetic state, the spins are aligned along *a*-axis; the nearest-neighbor (nn) spins are antiparallel along *a*- and *c*-, and parallel along shortest *b*-axis. The nesting ordering wave vector is $q$ = (101)$_O$ or (½ ½ 1)$_T$, relative to the tetragonal (T) or orthorhombic (O) nuclear cells.[60] In the top panels, the order parameter to the SDW order is seen by the intensity of the magnetic reflection (105)$_O$ / (½½5)$_T$; for tracking $T_S$, the intensity of the (400)$_O$ / (220)$_T$ nuclear peak was measured with warming. Similar to 122, the nuclear (220)$_T$ is expected to split to (400)$_O$ and (040)$_O$ orthorhombic Bragg reflections below $T_S$ in Ag-122. The increased intensity of structural peak is due to reduced extinction effect by the structural transition from tetragonal to orthorhombic lattice. The temperature-dependence of full peak width at half maximum (FWHM) of (400)$_O$ / (220)$_T$ is shown in the bottom panels of Fig. 5; the peak broadening indicates the splitting of (220)$_T$ peak. For Ag-122, we surprisingly find evidence that the magnetic and structural transitions occur roughly at the same temperature, i.e. $T_N = T_S$ = 125 K for $x$ = 0.05, $T_N = T_S$ = 84 K for $x$ = 0.029, and $T_N = T_S$ = 80 K for both $x$ = 0.035 and 0.045. Such behavior that $T_N$ and $T_S$ are coupled is different from behavior of Cu- or Au- dopants in 122,[37,50,61] or any other transition-metal doped systems.

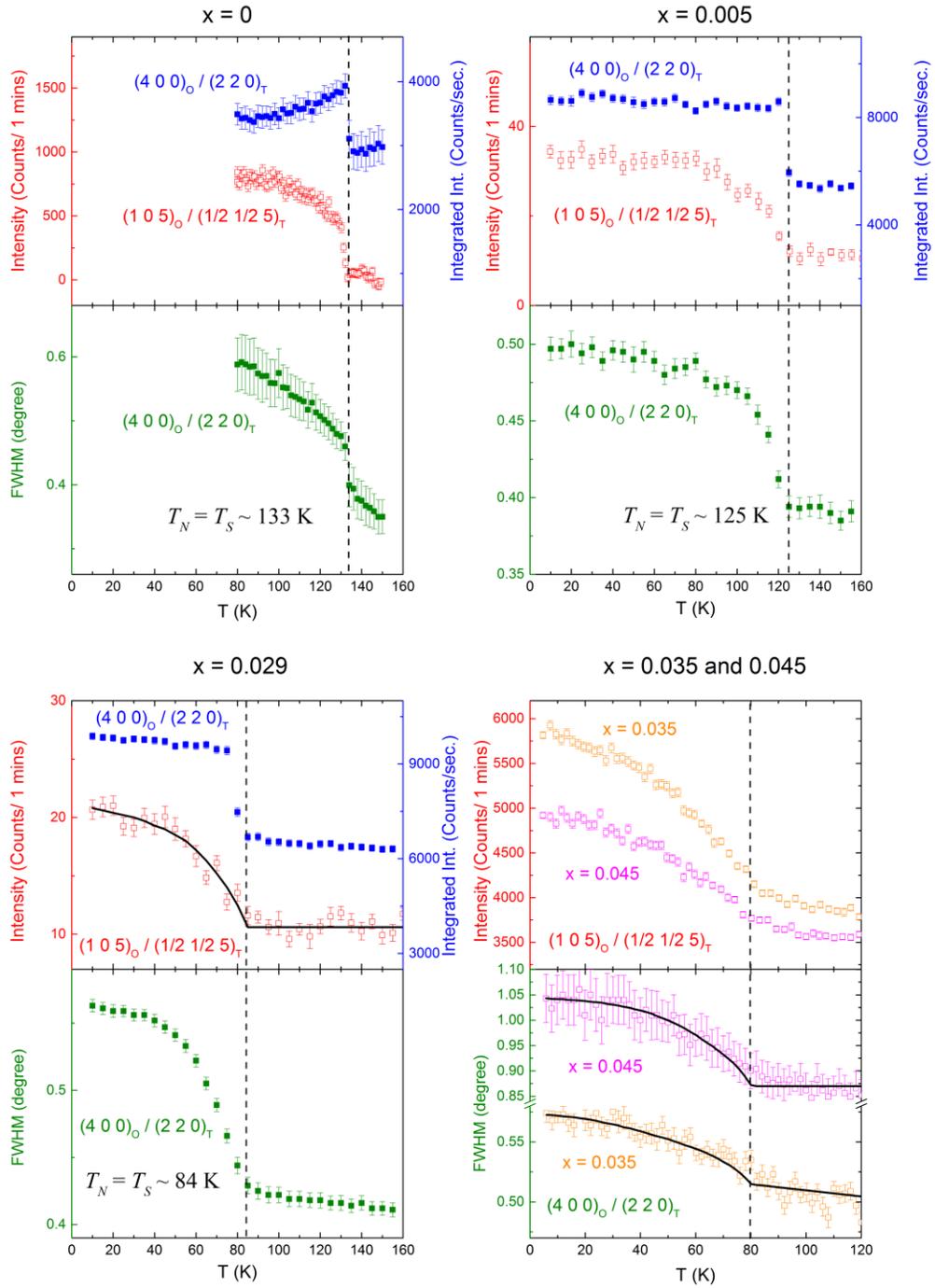

**Figure 5** For Ba(Fe$_{1-x}$Ag$_x$)$_2$As$_2$ crystals, neutron diffraction results for $x$ = 0, 0.005, 0.029, 0.035, and 0.045. The temperature dependence of Bragg reflection results upon warming. Top panels: integrated intensity of the nuclear peak (400)$_O$/(220)$_T$ and the peak intensity of the magnetic peak (105)$_O$/(½½5)$_T$; curved line is a guide for eyes. The solid/dashed line marks the structural/magnetic transition. Bottom panels: full peak width of (400)$_O$/(220)$_T$ at half peak maximum.

Based on the measurement results presented above, the $T$-$x$ phase diagram is constructed for Ag-122 system, shown in Fig. 6. Upon Ag doping, the structural and magnetic transition temperatures decrease monotonically. Unlike other electron-doped 122s, $T_N$ and $T_S$ values coincide for Ag-122 without splitting. This phase diagram is clearly divided into two regions: the tetragonal non-magnetic (TET/NM) and the orthorhombic antiferromagnetic (ORTH/AFM). The comparisons of $T_N$ vs. $x$ for Ag- and Cu-122 are illustrated in the inset of Fig. 6. For Cu-122, superconductivity with $T_C \sim 2$ K was found in $x = 0.044$ sample and $T_N$ suppressed linearly with $x$. For Ag-122 and for $x \leq 0.019$, $T_N$ follows the same guide line as Cu-122, while for $x > 0.019$, it starts to deviate approaching a potentially saturated value (of 78 K) around $x = 0.045$. This deviation in $T_N$ corresponds to those in $c$-lattice parameters and crystal features, and it may imply some change for the distance and interaction between FeAs layers that play an important role for persisting magnetism in Ag-122.

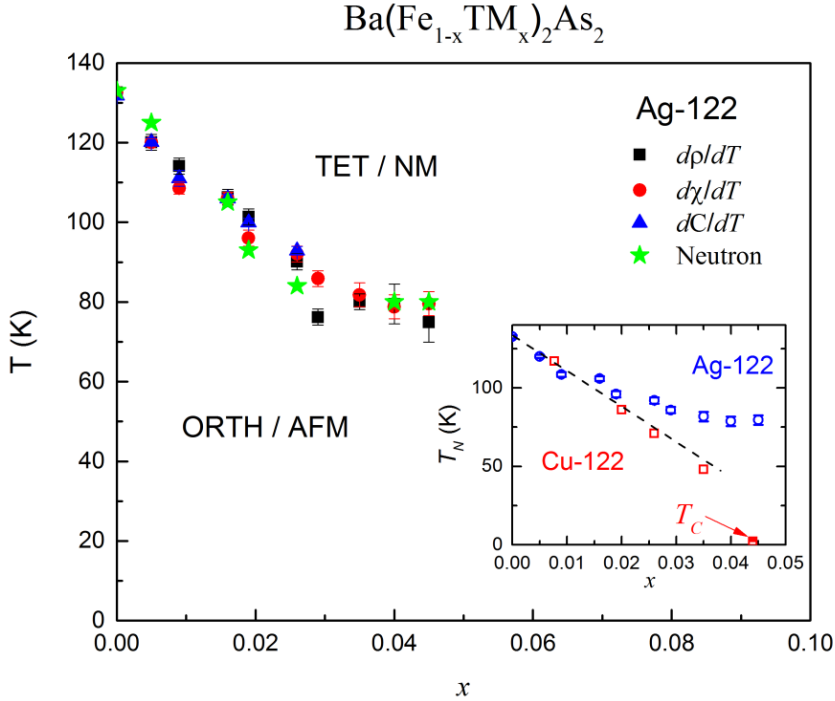

**Figure 6** T-$x$ phase diagram for Ag-122. Inset shows $T_N$ vs $x$ for Ag-122 and Cu-122 (data retrieved from ref.[37]).

### III. THEORETICAL CALCULATIONS

In an attempt to understand the observed behavior, in particular, the lack of superconductivity and maintenance of a magnetic state for Ag-122, we have performed first principles density functional theory calculations using the commercially available all-electron code WIEN2K[62] within the local density approximation (LDA). We have chosen the LDA rather than the GGA as it may give a better account of properties in 122s. There are two basic effects that must be considered in these calculations – the effect of *electron count* or charge doping, and the effects of *strain* – the change in structural parameters such as lattice constants with doping. In order to properly account for these effects we have performed calculations in the tetragonal unit cell with one of the four equivalent Fe atoms substituted by Ag, Cu, Co or Mn, using the room-temperature structural parameters for 122 doped with each of these at $x \sim 0.04$.

Note that the experimental observations for 122 with each of these dopants are very different: Ag doping produces *no* superconductivity, Cu a very slight superconductivity with maximum $T_c$ of 2 K, Co doping a substantial superconducting dome with $T_c$ values of 20 K, and Mn no superconductivity. What is the reason for these disparate behaviors? To answer this question, in this work we consider the impact of the dopant on electronic scattering. Such scattering has long been known to negatively impact superconductivity. One measure of such scattering, as described in [10], is the effective disruption of the electronic structure occasioned by the dopant atom. To the extent that the states arising from the dopant atom mirror those of the overall system, both in their energy range and magnitude, one may expect that such a dopant introduces charge without inordinate electronic scattering, and thus might induce superconductivity. On the other hand, a dopant that induces an energetically distinct set of electronic states can reasonably be expected to create large scattering and thus be less likely to induce (or allow) superconductivity. The simplest way to access this is through examination of the calculated density-of-states (DOS).

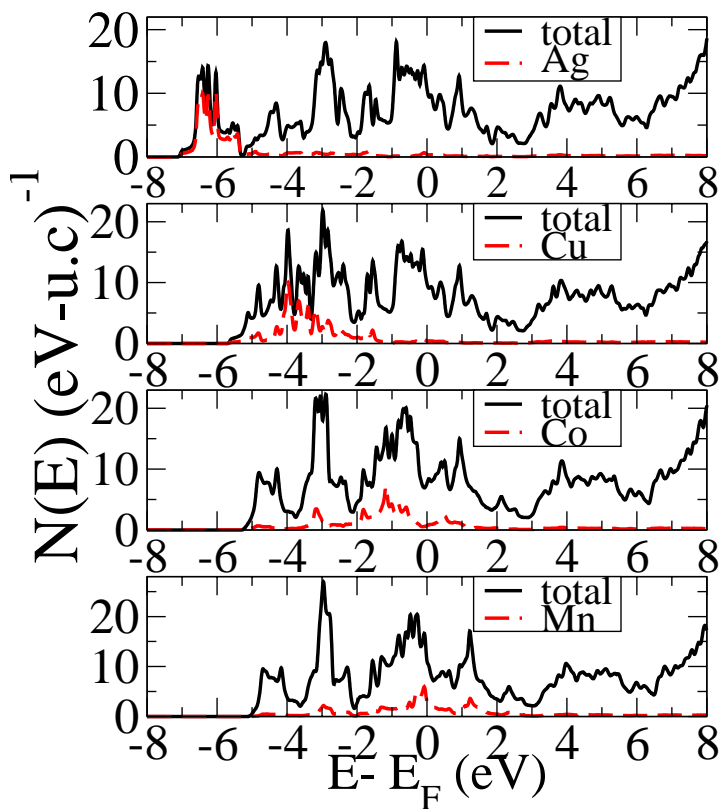

**Figure 7** The calculated densities-of-states for Ba-122 doped with each of the indicated dopants

We depict the results of these calculations in Figure 7. The top panel shows the result for Ag doping. We see that the Ag creates essentially a band separate from the rest of the electronic states, falling between 5.5 and 7.5 eV below the Fermi energy. This is a wholesale disruption of the electronic structure and is indicative of intense electronic scattering, consistent with the lack of superconductivity in Ag-122. It is instructive to compare the Ag results to the Cu results. In this latter case, the Cu states, unlike the Ag states, do not form a separate band, but fall broadly within a range 1.5 to 6 eV below $E_F$. While there is still substantial scattering associated with this band (it does not 'mirror' the overall DOS), it is likely to be significantly less than for Ag doping, suggesting the possibility of superconductivity for Cu doping. It is instructive to consider the last two results – for Co and Mn doping. The Co-doped DOS shows significantly less disruption than either Ag or Cu – the Co states fall within a wide range from -4 eV to +1 eV, and the maximum Co DOS at -1.1 eV is not far from the overall DOS maximum at -0.6 eV. In fact, Co-122 has larger superconducting dome ($T_{C,max} \approx 22$ K, and $\Delta x = 0.1$) than Cu-122 ($T_{C,max} = 2$ K, and $\Delta x = 0.015$). [37,57] Hence the electronic scattering is still less here, again generally consistent with the substantial superconducting dome for Co doping. Surprisingly, of the four dopants considered the dopant that produces the least disruption to the electronic structure is Mn. Here the DOS generally mirrors the overall electronic structure for a wide range around $E_F$. In this range its value is near the ¼ ratio of the total DOS that would indicate no disruption, and no separate band, as in the case of Ag, is visible. Experimentally Mn doping produces no superconductivity but rather yields a magnetic ground state, which is likely due to the tendency of Mn to retain a distinct local moment. An additional effect, not considered here, is magnetic scattering, which one would expect to be intense for Mn doping. However, this DOS result does suggest some possibility for achieving superconductivity via hole doping of the Fe site, which has never been achieved. Therefore, one possible explanation for the maintenance of magnetism in Ag-122 can be found in the Fermi-level densities-of-states for the four supercells – Ag, Cu, Co and Mn. Respectively these values (per Rydberg-unit cell) are 145.4, 133.4, 103.3 and 136.0, so that in fact the Fermi level DOS is *highest* for Ag doping. Given that magnetism in these families results partly from Stoner physics,[63] in which high density-of-states plays a key role, it is possible that details of the Ag-Fe interaction cause this interaction to be *less* disruptive to magnetism than for other dopants. We note that the Co-doped cell, which produces the largest superconducting dome, shows the lowest Fermi-level DOS, and so might be considered more effective at disrupting the magnetic state and allowing superconductivity to emerge. Further consideration of this topic will be in our future work.

In conclusion, this manuscript is the first study of silver substitution in $BaFe_2As_2$. We represent *T-x* phase diagram through extended experimental work. We describe the persistent magnetism in Ag-122 due to electronic scattering, and its relevant lattice parameter changes with *x* that are even reflected in crystalline morphologies.

## ACKNOWLEDGMENTS

This work was primarily supported by the U. S. Department of Energy (DOE), Office of Science, Basic Energy Sciences, Materials Science and Engineering Division (LL, DP, AS). The work at ORNL's HFIR (HC) was sponsored by the Scientific User Facilities Division, Office of Basic Energy Sciences, U. S. Department of Energy. S.K. would like to acknowledge DOE Office of Science Graduate Student Research Program award for funding under contract number DE-AC05-06OR23100.